%% file: main.tex
\documentclass[sigconf]{acmart}

\usepackage{xcolor}
\usepackage{xspace}
\usepackage{multirow}
\usepackage{booktabs}
\usepackage{enumitem}
\usepackage[normalem]{ulem}
\usepackage{bm}

\usepackage{algorithm}
\usepackage{algpseudocode}
\useunder{\uline}{\ul}{}

\usepackage[normalem]{ulem}
\usepackage{amsmath}
\newcommand{\stkout}[1]{\ifmmode\text{\sout{\ensuremath{#1}}}\else\sout{#1}\fi}
\newcommand{\Zstroke}{%
  \stkout{Z}%
}

\newcommand{\hide}[1]{}

\newcommand{\ours}{DARec\xspace}
\newcommand{\oursbase}{DARec$_{\text{base}}$\xspace}

\setcopyright{acmlicensed}
\copyrightyear{2018}
\acmYear{2018}
\acmDOI{XXXXXXX.XXXXXXX}

\acmConference[Conference acronym 'XX]{Make sure to enter the correct
  conference title from your rights confirmation emai}{June 03--05,
  2018}{Woodstock, NY}
\acmISBN{978-1-4503-XXXX-X/18/06}

\begin{document}

\title{Beyond Inter-Item Relations: Dynamic Adaption for Enhancing LLM-Based Sequential Recommendation}

\author{CanYi Liu}
\email{liucanyi01@stu.xmu.edu.cn}
\affiliation{
  \institution{Key Laboratory of Multimedia Trusted Perception and Efficient Computing, Ministry of Education of China}
  \country{Xiamen University, Xiamen, Fujian, China}
}
\author{Wei Li}
\email{liwei2002@stu.hit.edu.cn}
\affiliation{
  \institution{Faculty of Computing, Harbin Institute of Technology}
  \city{Harbin}
  \state{Heilongjiang}
  \country{China}
}

\author{Youchen (Victor) Zhang}
\email{vzhang996@g.ucla.edu}
\affiliation{
  \institution{University of California, Los Angeles}
  \city{Los Angeles}
  \state{California}
  \country{United States}
}

\author{Hui Li, Rongrong Ji}
\email{{hui, rrji}@xmu.edu.cn}
\affiliation{
  \institution{Key Laboratory of Multimedia Trusted Perception and Efficient Computing, Ministry of Education of China}
  \country{Xiamen University, Xiamen, Fujian, China}
}

\begin{abstract}
Sequential recommender systems (SRS) predict the next items that users may prefer based on user historical interaction sequences. Inspired by the rise of large language models (LLMs) in various AI applications, there is a surge of work on LLM-based SRS. Despite their attractive performance, existing LLM-based SRS still exhibit some limitations, including neglecting intra-item relations, ignoring long-term collaborative knowledge and using inflexible architecture designs for adaption. To alleviate these issues, we propose an LLM-based sequential recommendation model named \ours. Built on top of coarse-grained adaption for capturing inter-item relations, \ours is further enhanced with (1) context masking that models intra-item relations to help LLM better understand token and item semantics in the context of SRS, (2) collaborative knowledge injection that helps LLM incorporate long-term collaborative knowledge, and (3) a dynamic adaption mechanism that uses Bayesian optimization to flexibly choose layer-wise adapter architectures in order to better incorporate different sequential information. Extensive experiments demonstrate that \ours can effectively handle sequential recommendation in a dynamic and adaptive manner.
\end{abstract}



\maketitle

\input{tex/introduction}   
\input{tex/method}
\input{tex/experiment}

\input{tex/related_work}
\input{tex/conclusion}

\bibliographystyle{ACM-Reference-Format}
\bibliography{main}

\input{tex/appendix}

\end{document}

%% file: tex/introduction.tex

\section{Introduction}
\label{sec:intro}

Sequential recommender systems (SRS) suggest items that may interest users by modeling user historical interaction sequences.
Aiming at helping users find items more efficiently and online service providers better promote items, SRS have been applied in a wide variety of online applications including but not limited to online shopping (e.g., Amazon), music listening (e.g., Spotify) and news reading (e.g., Yahoo News)~\cite{FangZSG20}. 

There has been a tremendous amount of work on sequential recommendation in the last decades~\cite{WangHWCSO19,WangZ0ZW022,QuadranaCJ18}.
In particular, deep learning techniques have greatly boosted the prosperous development of SRS~\cite{FangZSG20}.
Nevertheless, deep learning-based SRS still face various challenges.
One representative issue is that, due to model scale and data size~\cite{zhao2024recommendersystemseralarge}, the understanding of deep learning-based SRS is limited to the context of SRS (e.g., the descriptions of items provided by the online service), making it difficult for them to think like humans with broad world knowledge.

Thanks to the rise of large language models (LLMs)~\cite{zhao2023survey}, the aforementioned issue can be alleviated by introducing LLMs into SRS.
Firstly, sequential recommendation data shares some similarities with natural language text in that they are both sequential and represented by individual items/tokens.
It is natural to consider modeling sequential recommendation in a way similar to language modeling and leverage LLMs to capture complex sequential item relations.
Moreover, items in contemporary SRS are typically accompanied by textual side information (e.g., item titles, attributes and descriptions) that can be better understood by LLMs with world knowledge beyond the context of SRS.
Therefore, there is a surge of work on applying LLMs in SRS~\cite{zhao2024recommendersystemseralarge,abs-2402-11143,0025ZG23,LinDX24}.

Despite the promise of LLM-based SRS, we argue that existing works still exhibit some limitations:
    
\vspace{3pt}
\noindent\textbf{P1: Neglect Intra-Item Relations}: Sequential relations are under-explored, resulting in sub-optimal performance. Most LLM-based SRS follow the traditional SRS paradigm that captures \emph{inter-item} relations in user historical interaction sequences (i.e., item level), neglecting that the minimal modeling granularity in LLM-based SRS is a token in item textual data. Hence, existing methods are coarse-grained and overlook the \emph{intra-item} relations (i.e., relations of tokens in an item's textual data).

\vspace{3pt}
\noindent\textbf{P2: Ignore Long-Term Collaborative Knowledge}: Previous LLM-based SRS focus on sequential item information in order to fit the sequential modeling nature of LLMs. However, they neglect long-term collaborative knowledge (i.e., user/item representations that can be captured by traditional collaborative filtering (CF) methods). Although user preferences and item properties can be extracted from historical interaction sequences, it is well acknowledged that LLM is not good at modeling long-sequence inputs~\cite{abs-2401-01325} and cannot model such long-term collaborative knowledge well.

\vspace{3pt}
\noindent\textbf{P3: Inflexible Adaption}: To reduce the computation cost, most LLM-based SRS opt for parameter-efficient fine-tuning (PEFT)~\cite{abs-2403-14608} to adapt general LLMs for sequential recommendation by fine-tuning only a small amount of parameters. Inserting small adapters is one representative PEFT approach for LLM-based SRS and it generally exhibits better performance than other PEFT methods~\cite{FuY0YCCZWP24}. However, the design of adapters in existing LLM-based SRS is typically pre-defined, leaving the adaption inflexible on different data.

To address the above issues, we propose an LLM-based SRS that uses dynamically adapted LLM to incorporate intra-item information and collaborative knowledge, in addition to inter-item relation captured by existing works, into sequential recommendation. 
We name it \ours where DA stands for dynamic adaption.
Note that we focus on using adapter, as it generally exhibits better performance than other PEFT methods on recommendation tasks~\cite{FuY0YCCZWP24}.
However, our design is orthogonal to other PEFT approaches like LoRA~\cite{HuSWALWWC22} and has the potential to be combined with other methods.
In summary, the contributions of this work are:
\begin{enumerate}[leftmargin=15pt,topsep=1pt,itemsep=0.5pt]
    \item We design several coarse-grained adaption tasks, including semantic alignment, linking next item and distinguishing hard samples, to adapt LLM for sequential recommendation. Based on these tasks, \oursbase which uses the standard adapter design for LLM-based SRS and is the basic version of \ours, can capture coarse-grained inter-item relations for sequential recommendation.
    
    \item To overcome \textbf{P1}, we propose context masking that captures relations of tokens within the textual information of an item. Context masking is used together with causal masking, the standard attention mechanism of LLM, as two adapter modules to help LLM better understand token and item semantics in the context of SRS.

    \item To address \textbf{P2}, we design collaborative knowledge injection that jointly trains a small CF model and LLM-based SRS. The learned user/item representations from the CF model are injected into attention computation so that LLM-based SRS can benefit from long-term collaborative knowledge.

    \item To solve \textbf{P3}, we leverage Bayesian optimization with reparameterization to search binary, discrete parameters for dynamically deciding the suitable adapter design for each LLM layer, bringing flexibility to \ours.

    \item Based on \oursbase and our designed techniques in (2), (3) and (4), we propose DARec, a novel LLM-based sequential recommendation model designed to enhance LLM-based sequential recommendation by extending beyond inter-item relationship modeling. We conduct extensive experiments and the results demonstrate that \ours can effectively handle sequential recommendation in a dynamic and adaptive manner.

\end{enumerate}

%% file: tex/method.tex

\hide{

\section{Problem Formulation}
\label{sec:Problem Formulation}
A typical recommendation scenario generally includes a set of users $ U $ and a set of items $ I $. 
For each user $ u \in U $, we have a sequence $ S_u = \left[ i_1, i_2, \dots , i_n \right] $, where the interactions are ordered chronologically, and $ i_j \in I $ represents an interaction between user $ u $ and item $ i_j $. 
Additionally, we have a nested dictionary $D$. Each item $i_j \in I$ has a unique corresponding text attribute dictionary $D_{ID_{i_j}}=\left\{(k_1: v_1),\dots,(k_m: v_m)\right\}$, where $ID_{i_j}$ is the ID of item $i_j$, $k_1,\dots,k_m$ are the keys of the text attributes of item $i_j$, and $v_1,\dots,v_m$ are the values of these text attributes. 

For convenience, in the following sections, we will use $i_j$ to denote the ID of item $i_j$. Similarly, $u_j$ will denote the user's ID $u_j$.

We aim to construct a model $M$ that predicts the next item $ i_{n+1} $ that user $ u $ will interact with by learning from the user's historical interaction sequence $ S_u $ and the items' textual attributes $ T_i $. 
It can be formalized as modeling the probability over all possible items for user $u$ at time step $n+1$:

\begin{displaymath}
  i_{n+1} = \arg\max_{i \in I} p \big(i \mid S_u, D, M\big)
\end{displaymath}

To model the sequential recommendation task, we construct sequential behavior data containing the interaction sequence $ S_u = \left[ i_1, i_2, \dots, i_n \right] $ for each user $u$, where items in each sequence are in chronological order.

}

\begin{figure}[t]
  \centering
  \includegraphics[width=0.85\columnwidth]{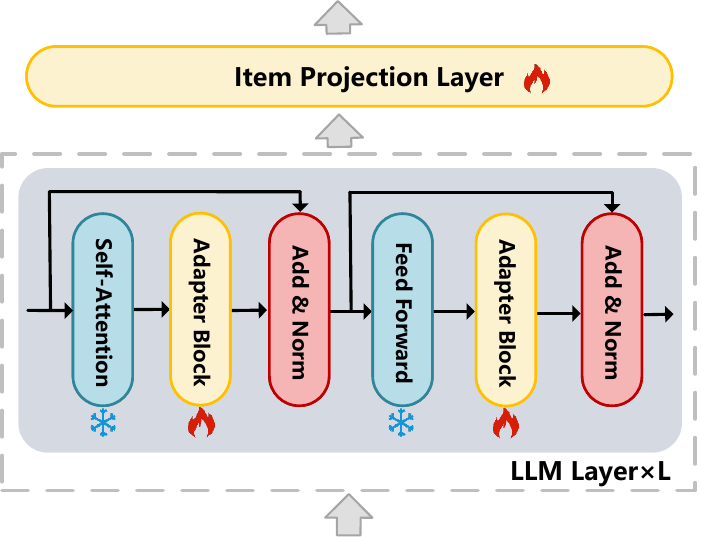}
  \vspace{-10pt}
  \caption{Overview of \oursbase. The figure uses the architecture of the standard Transformer layer while the actual design for ``Add \& Nrom'' varies according to detailed LLM.}
  \vspace{-10pt}
  \label{fig:base}
\end{figure}

\section{Our Method \ours}

\subsection{Overview}

\oursbase, the basic version of \ours, consists of an LLM and an item projection layer.
We adopt the standard design for adapters~\cite{FuY0YCCZWP24} in \oursbase. 
As shown in Fig.~\ref{fig:base}, \oursbase has two adapter blocks inserted into each LLM layer for PEFT, and an item projection layer for mapping the LLM space to the item space.
We design various coarse-grained adaption tasks that capture inter-item relations to endow \oursbase with the ability to adapt LLM for sequential recommendation (Sec.~\ref{sec:adapt}).

Based on \oursbase, we enhance \ours with context masking, collaborative knowledge injection and a dynamic adaption mechanism (Sec.~\ref{sec:enhance}) to further incorporate intra-item and long-term collaborative knowledge in a \emph{dynamic} and \emph{adaptive} manner.

\subsection{Adapt LLM for Sequential Recommendation}
\label{sec:adapt}

We first propose several coarse-grained adaption tasks to help \oursbase capture inter-item relations and adapt to sequential recommendation.

\begin{figure*}[t]
  \centering
  \includegraphics[width=\textwidth]{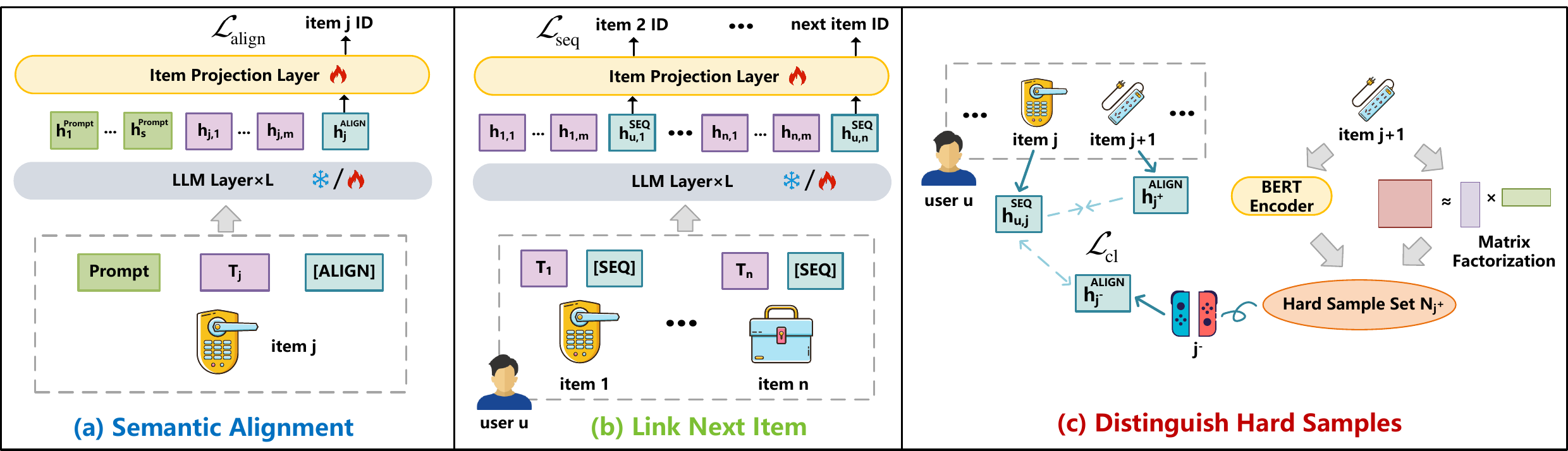}
  \vspace{-20pt}
  \caption{Adapt LLM for Sequential Recommendation (\oursbase): (1) $\mathbf{h}_{*}^{\textbf{Prompt}}$ indicates the hidden state for a token in the prompt. (2) $\mathbf{h}_{j,*}$ is the hidden state for a token in the item ``sentence'' of item $j$. (3) $\mathbf{h}_{j}^{\textbf{ALIGN}}$ is the hidden state for [ALIGN] of item $j$. (4) $\mathbf{h}_{u,j}^{\textbf{SEQ}}$ is the hidden state for [SEQ] of item $j$ in user $u$'s historical interaction sequence.}
  \vspace{-5pt}
  \label{fig:adapt}
\end{figure*}

\subsubsection{Item Representation}
For each item $ i $, we have its corresponding attribute dictionary $ {Dic}_i=\{k,v\} $, where $ k $ and $ v $ consist of words.
Following RecFormer~\cite{li2023text}, we flatten $ {Dic}_i $ into an item ``sentence'': $ T_i = \{ k_1, v_1, \dots, k_g, v_g \} $ as the item representation for $i$ where $g$ is the number of key-value pairs. 
This design has been prevalently applied in LLM-based SRS~\cite{HuXZFWHLTZ24,RenCYLJCZM024,ZhengCQZ024}.

\subsubsection{Semantic Alignment}
\label{sec:alignment}

Items are typically represented by IDs (e.g., ``X2112'') in SRS. 
However, these item IDs are rare in the training corpora of LLM.
Hence, given the recommendation context, it is difficult for LLM to understand item IDs. 
To enhance the understanding of LLM on item IDs, as shown in Fig.~\ref{fig:adapt}(a), we conduct semantic alignment between a special token $\left [ \text{ALIGN} \right ] $ and the corresponding item ID.

Specially, for an item $i$, we concatenate a prompt, the item ``sentence'', and a special token $\left [ \text{ALIGN} \right ] $, and feed \{Prompt, $T_i$, $\left [ \text{ALIGN} \right ] $\} to LLM.
Here, we use the prompt:``Give the ID of the item described by:...''.
Then, we adopt an item projection layer (a single-layer feedforward network) to convert $\mathbf{h}^{\text{ALIGN}}_{i}$, the hidden state corresponds to $\left [ \text{ALIGN} \right ] $ from the last layer in LLM, to the item probability distribution.
$\mathbf{h}^{\text{ALIGN}}_{i}$ is regarded as the learned representation of $i$, similar to using the representation of a special token [CLS] to represent a sentence when using a pre-trained model in the text classification task~\cite{DevlinCLT19,abs-1907-11692}.
During optimization, we use cross-entropy loss between the item probability distributions and actual item IDs to align hidden states and corresponding item IDs:
\begin{equation}
\small 
    \hat{\mathbf{y}}_{i} = \text{Softmax}(\textbf{W}_{\text{item}} \cdot \mathbf{h}^{\text{ALIGN}}_{i} + \textbf{b}_{\text{item}}),\,\,\mathcal{L}_{\text{align}} = - \sum_{i}  y_{i} \log \hat{\mathbf{y}}_{i} \nonumber
\end{equation}
where $\textbf{W}_{\text{item}}$ and $\textbf{b}_{\text{item}}$ are learnable parameters, and $y_{i}$ is the actual item ID of item $i$.

Through semantic alignment, \oursbase can capture the relationship between item IDs and the recommendation context where item IDs appear, thereby gaining a better understanding of items.

\subsubsection{Link Next Item}
\label{sec:predict_next}

We further endow LLM with the ability to predict next items.
Each item $i$ in the historical interaction sequence $S_u$ of the user $u$ can be denoted by its item ``sentence''. 
We append a special token $\left [ \text{SEQ} \right ] $ to the end of each item ``sentence'' to separate items and construct the sequential textual data for $u$:
\begin{equation}
\small
    X_{u} = \left \{ T_{1}, \left [ \text{SEQ} \right ], \dots, T_{j}, \left [ \text{SEQ} \right ], \cdots, T_{n}, \left [ \text{SEQ} \right ] \right \},\nonumber
\end{equation}
where $n$ is the number of items in $u$'s historical interaction sequence.

As depicted in Fig.~\ref{fig:adapt}(b), $X_u$ is then fed into LLM and \oursbase uses the item projection layer to map the hidden state $\mathbf{h}^{\text{SEQ}}_{u,i}$ from the last LLM layer that corresponds to the current item $i$ to the next item ID.
We use cross-entropy loss between predictions and actual next-item IDs during optimization:
\begin{equation}
\small
    \hat{\mathbf{z}}_{u,i+1} = \text{Softmax}(\textbf{W}_{\text{item}} \cdot \mathbf{h}^{\text{SEQ}}_{u,i} + \textbf{b}_{\text{item}}),\,\,\mathcal{L}_{\text{seq}} = - \sum_{u}\sum_{j=2}^{n}  z_{u,j} \log \hat{\mathbf{z}}_{u,j}  \nonumber
\end{equation}
where $z_{u,j}$ is the actual ID of the $j$-th item in $u$'s historical interaction sequence.

\subsubsection{Distinguish Hard Samples}
\label{sec:cl}

LLM is trained over vast amounts of text data but it does not see much sequential recommendation data.
Hence, LLM faces difficulty in distinguishing items in the sequential recommendation task.
As shown in Fig.~\ref{fig:adapt}(c), we adopt contrastive learning to enhance the ability of \oursbase to distinguish different items without requiring additional data.
Particularly, we focus on improving distinguishing \emph{hard samples}\footnote{The details of generating hard samples are provided in Appendix~\ref{sec:app_hard_sample}}.
For an item $j$, hard samples $N_j$ are items similar to $j$, but they do not co-occur with $j$ in any user historical interaction sequence $S$.

We randomly sample $n_{s}$ items from each $S$, excluding the item in the last position.
For a sampled item $j \in S_u$, \oursbase treats its subsequent item $j^{+}$ in $S_u$ as the positive item and randomly chooses $n_{cl}$ items from $N_{j^{+}}$ as the negative items $j^{-}$ (all $j^{-}$ forms the set $J^{-}$).
Recall that $\mathbf{h}^{\text{SEQ}}_{u,j}$ is correlated with the next-item ID $j^{+}$ (Sec.~\ref{sec:predict_next}), while $\mathbf{h}^{\text{ALIGN}}_{j^{+}}$ and $\mathbf{h}^{\text{ALIGN}}_{j^{-}}$ are aligned with IDs of $j^{+}$ and $j^{-}$ (Sec.~\ref{sec:alignment}), respectively.
\oursbase conducts contrastive learning using InfoNCE loss~\cite{abs-1807-03748} among $\mathbf{h}^{\text{SEQ}}_{u,j}$, $\mathbf{h}^{\text{ALIGN}}_{j^{+}}$ and $\mathbf{h}^{\text{ALIGN}}_{j^{-}}$ to enhance its understanding of intrinsic differences of items: $\mathbf{h}^{\text{SEQ}}_{u,j}$ is trained to be close to $\mathbf{h}^{\text{ALIGN}}_{j^{+}}$ and far away from $\mathbf{h}^{\text{ALIGN}}_{j^{-}}$:
\begin{equation}
\small
\begin{aligned}
    & f(\mathbf{h},\mathbf{\mathbf{h}^{'}}) = exp\left( \text{sim}(\mathbf{h}, \mathbf{\mathbf{h}^{'}})/\tau \right) \\
    \mathcal{L}_{\text{cl}} = - \sum_{S_u} \sum_{j=1}^{n_s} &\log \frac{f(\mathbf{h}^{\text{SEQ}}_{u,j}, \mathbf{h}^{\text{ALIGN}}_{j^+})}{f(\mathbf{h}^{\text{SEQ}}_{u,j}, \mathbf{h}^{\text{ALIGN}}_{j^+}) + \sum_{j^{-}\in J^{-}}f(\mathbf{h}^{\text{SEQ}}_{u,j}, \mathbf{h}^{\text{ALIGN}}_{j^-})},\nonumber
\end{aligned}
\end{equation}
where $\tau$ is the temperature parameter for scaling similarities and $sim(\cdot)$ is the cosine similarity. 

\begin{figure*}[t]
\centering
\begin{minipage}{.3\textwidth}
  \centering
  \includegraphics[width=\linewidth]{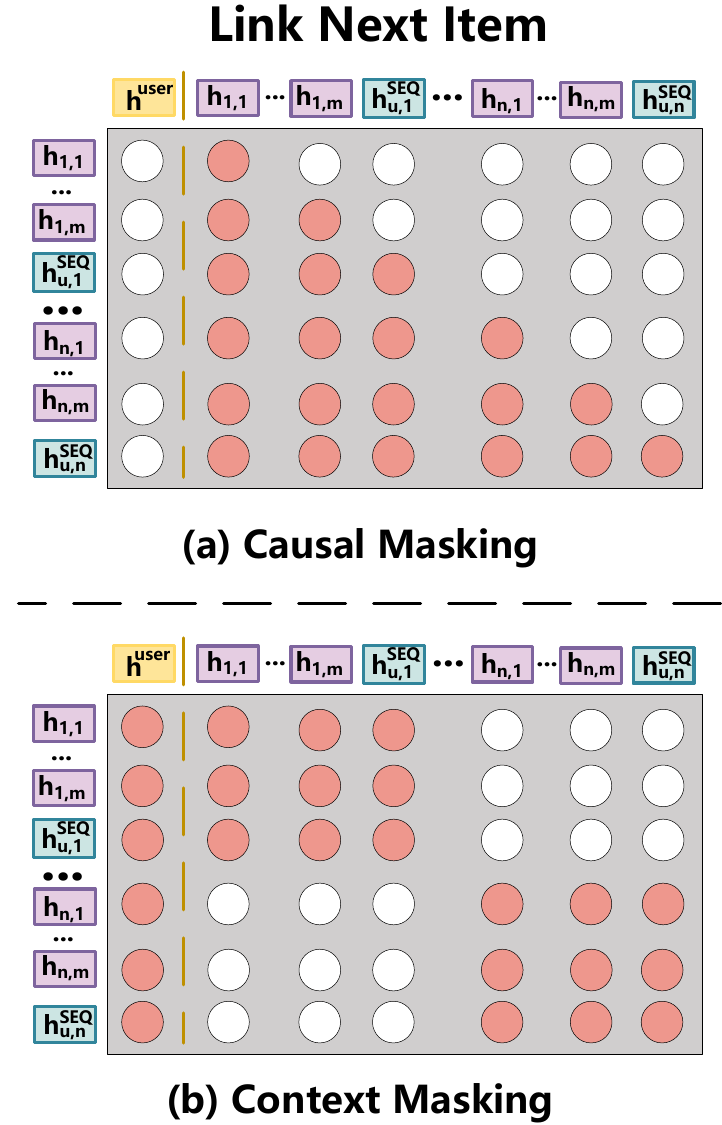}
  \vspace{-20pt}
  \captionof{figure}{Two Masking Mechanisms.}
  \vspace{-5pt}
  \label{fig:mask}
\end{minipage}%
\hspace{.02\textwidth}
\begin{minipage}{.67\textwidth}
  \centering
  \includegraphics[width=\linewidth]{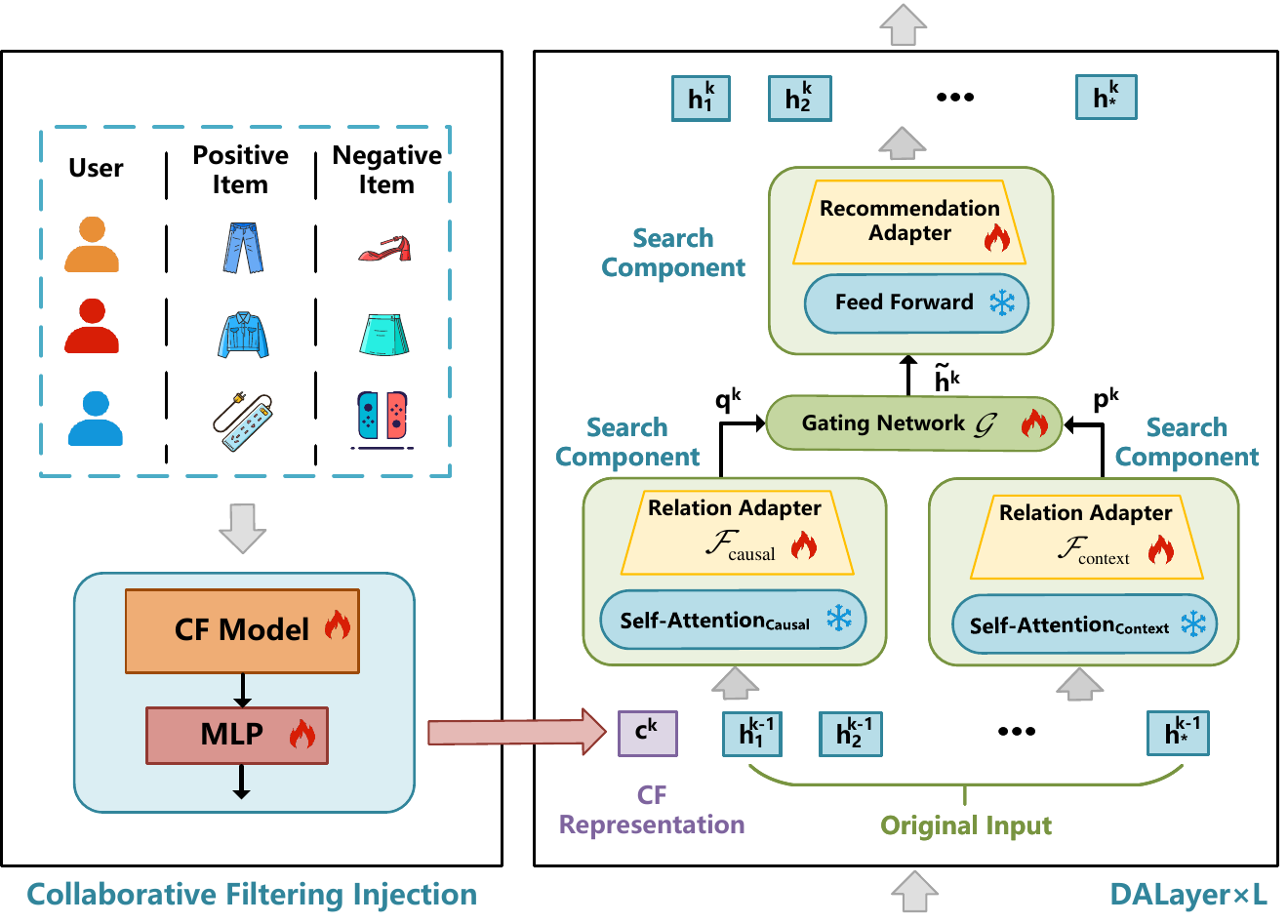}
  \vspace{-20pt}
  \captionof{figure}{Overview of Dynamic Adaptation in \ours.}             
  \vspace{-5pt}
  \label{fig:DAMoE}
\end{minipage}
\end{figure*}

\subsubsection{Recommendation Adapters}
\label{sec:rec_ada}
To reduce the computation overhead for fine-tuning \oursbase, as shown in Fig.~\ref{fig:base}, we keep the parameters of the feedforward network and the self-attention module in each LLM layer intact and fine-tune two injected recommendation adapters (a two-layer feedforward network) following the design of standard recommendation adapters~\cite{FuY0YCCZWP24}:
\begin{equation}
    \small
    \text{Adapter}(\mathbf{h}) =  \mathbf{W}_2\big(\mathbf{W}_1\cdot \mathbf{h} + \mathbf{b}_1) + \mathbf{b}_2 + \mathbf{h},\nonumber
\end{equation}
where $\mathbf{W}$ and $\mathbf{b}$ are learnable parameters.

\subsection{Incorporate Intra-Item Relation and Long-Term Collaborative Knowledge}
\label{sec:enhance}
In this section, we will illustrate how \ours, based on \oursbase, can capture intra-item relation and collaborative knowledge under the design of dynamic adaption. 

\subsubsection{Context Masking}

Coarse-grained adaption tasks only capture inter-item relations and do not emphasize on the relations between tokens within the same item (intra-item relations).
The reason is that standard decoder-only LLMs adopt causal masking~\cite{abs-2402-04779} to compute attention, where each token attends to prior and current tokens only, ignoring future tokens. 
As shown in Fig.~\ref{fig:mask}(a), causal masking does not set boundaries to separate tokens from different items.
\ours represents each item with multiple tokens (i.e., item ``sentence''), and using causal masking only will fail to fully model token semantics in the item context. 

We introduce context masking to address the above issue.
Fig.~\ref{fig:mask} depicts context masking and causal masking for linking next item (Sec.~\ref{sec:predict_next})\footnote{Fig.~\ref{fig:mask_align} in Appendix~\ref{sec:app_align} demonstrates semantic alignment (Sec.~\ref{sec:alignment}).}.
Context masking allows each token within an item ``sentence'' to attend only to other tokens within the same item (and collaborative representations introduced in the next section), disregarding tokens from other items. 
It is bi-directional but it focuses on intra-item relations. 
Mathematically, the definitions for causal masking and context masking are as follows:
\begin{equation}
\begin{small}
\begin{aligned}
    \textbf{MASK}_{\text{causal}}(i,j) &= \begin{cases}
        \text{True}, & \text{if } i \leq j \text{ and } j \neq 1 \\
        \text{False}, & \text{otherwise}
    \end{cases} \\
    \textbf{MASK}_{\text{context}}(i,j) &= \begin{cases}
        \text{True}, & \text{if } \text{SameItem}(i, j) \\
        \text{False}, & \text{otherwise}
    \end{cases}
\end{aligned}
\nonumber
\end{small}
\end{equation}
where $\textbf{MASK}_{\text{causal}}$ and $\textbf{MASK}_{\text{context}}$ are masking matrices of size $m \times (m+1)$ and $m$ is the maximum number of tokens in an item ``sentence'. 
The function $\text{SameItem}(i, j)$ returns $\text{True}$ if tokens $i$ and $j$ belong to the same item (or either of it is the collaborative representation), and $\text{False}$ otherwise.

\subsubsection{Collaborative Knowledge Injection}

LLM serves as an effective sequential model in \oursbase.
However, the long-term user/item collaborative knowledge is beneficial to SRS~\cite{abs-2404-11343} and LLM may not capture them well from long inputs.

We conduct collaborative knowledge injection and incorporate user preferences and item properties learned from a CF model $M_{cf}$ into the attention calculation of context masking, using them as cues to enhance \ours's learning.
$M_{cf}$ can be any embedding-based recommendation model trained over user-item interaction matrix using standard loss function $\mathcal{L}_{cf}$ (e.g., Bayesian Personalized Ranking) and it encodes each user $u$ and item $i$ into the representations $\mathbf{e}_u^{user}$ and $\mathbf{e}_i^{item}$, respectively.

\ours first uses a single-layer feedforward network to project $\mathbf{e}_u^{user}$ and $\mathbf{e}_i^{item}$ to new representation spaces $\mathbf{h}^{user}_u$ and $\mathbf{h}^{item}_i$ with the same dimension as LLM's hidden state, respectively.
Then, inspired by prefix tuning~\cite{LiL20}, we concatenate the projected collaborative representations ($\mathbf{h}^{user}$ or $\mathbf{h}^{item}$, i.e., $\mathbf{c}^k$ in Fig.~\ref{fig:DAMoE}) with hidden states of inputs to each LLM layer.

\hide{
    For semantic alignment for an item $j$ (Sec.~\ref{sec:alignment}), the input becomes:
    \begin{equation}
    \small
    \{ \mathbf{h}^{item}_j, \text{Prompt}, T_j, [\text{ALIGN}]\}.\nonumber
    \end{equation}
    For predicting next item for a user $u$ (Sec.~\ref{sec:predict_next}), the input becomes:
    \begin{equation}
    \small
        X_{u} = \left \{ \mathbf{h}^{user}_u, T_{1}, \left [ \text{SEQ} \right ], \dots, T_{j}, \left [ \text{SEQ} \right ], \cdots, T_{n}, \left [ \text{SEQ} \right ] \right \}.\nonumber
    \end{equation}
}

\subsubsection{Dynamic Adaption}
\label{sec:bo}

For the $k$-th layer in LLM, \ours derives two hidden states $\mathbf{q}^{k}$ and $\mathbf{p}^{k}$ for the input $\mathbf{h}^{k-1}$ using causal masking $\textbf{MASK}_{\text{causal}}$ and context masking $\textbf{MASK}_{\text{context}}$, respectively:
\begin{equation}
\small
\begin{aligned}
    \mathbf{q}^{k} &= \text{Self-Attention}(\mathbf{h}^{k-1}, \textbf{MASK}_{\text{causal}})\\
    \mathbf{p}^{k} &= \text{Self-Attention}(\mathbf{h}^{k-1}, \textbf{MASK}_{\text{context}})\nonumber
\end{aligned}
\end{equation}
where $\mathbf{h}^{k-1}$ is the hidden state of a token from the $(k-1)$-th layer\footnote{Appendix~\ref{sec:app_att} provides the definition of $\text{Self-Attention}(\mathbf{h}, \textbf{MASK})$}.

\ours adopts two relation adapters and each is responsible for either causal masking or context masking.
Then, we design a dynamic adaption mechanism that uses a router to dynamically compose two adapter modules.
The router computes the probability of input being sent to each adapter.
Original LLM layers in \oursbase are replaced with dynamically adapted layers (DALayer).
Fig.~\ref{fig:DAMoE} depicts the design of DALayer.
Each DALayer consists of two relation adapters $\mathcal{F}_{\text{causal}}$ and $\mathcal{F}_{\text{context}}$, a router $\mathcal{G}$, and one recommendation adapter.
The key of dynamic adaption is the design of the layer-wise dynamic selection of parameter-efficient relation adapters and the router.

\vspace{5pt}
\noindent\textbf{Layer-Wise Dynamic Selection of Relation Adapters.} 
To reduce the computation demands, 
we freeze feedforward networks and attention modules, and only fine-tune the relation adapters ($\mathcal{F}_{\text{causal}}$ and $\mathcal{F}_{\text{context}}$) and the router ($\mathcal{G}$). 

There are various choices for each adapter module.
Enforcing a unified design for adapters in all DALayers is inflexible and will bring sub-optimal performance as shown in our experiments (Sec.~\ref{sec:contribution}).
However, manually tuning the architecture of each adapter is a time-consuming and human-intensive process.
Hence, we propose a dynamic selection method to automatically find the appropriate design of adapters in each DALayer.
In \ours, we focus on choosing between serial adapter and parallel adapter\footnote{Fig.~\ref{fig:two_adapter} in Appendix~\ref{sec:app_adapter} provides their visualization.} for $\mathcal{F}_{\text{causal}}$, $\mathcal{F}_{\text{context}}$ and the recommendation adapter since serial adapter and parallel adapter are most representative adapter designs~\cite{abs-2403-14608}:
\begin{enumerate}[leftmargin=15pt,topsep=1pt,itemsep=0.5pt]
    \item \textbf{Serial Adapter:} A two-layer feedforward network is positioned between the attention module and the gating network: 
    \begin{equation}
    \small
    \text{Adapter}_{\text{Serial}}(\mathbf{h})=\text{Adapter}\big(\text{Self-Attention}(\mathbf{h}, \textbf{MASK})\big).\nonumber
    \end{equation}
    
    \item \textbf{Parallel Adapter:} A two-layer feedforward network runs alongside the attention module, i.e., it reorganizes the adapter as a parallel side network:
    \begin{equation}
    \small
    \text{Adapter}_{\text{Parallel}} (\mathbf{h})=  \Big(\text{Self-Attention}(\mathbf{h}, \textbf{MASK}) + \text{Adapter}(\mathbf{h})\Big)/2.\nonumber
    \end{equation}
\end{enumerate}

In each part, either a serial adapter or a parallel adapter can be applied.
Given $L$ DALayers, there are $2^{3L}$ possible designs.
Hence, the search process for each DALayer becomes tuning $3L$ binary, discrete parameters in the discrete space $\Zstroke=\Zstroke^{(1)}\times \cdots \times \Zstroke^{(3L)}$.
In the following, we use $\mathbf{z}$ to denote the vector $(z^{(1)},
\cdots,z^{(3L)})$ of which each element is a binary parameter.

Bayesian optimization (BO)~\cite{WangJSO23} is a popular method for black-box optimization that can be applied in this search task.
BO uses a probabilistic surrogate model of the objective function $f$ (we use Mean Reciprocal Rank) and an acquisition function (AF) that provides utility values for evaluation during optimization.
The maximizer of AF is selected as the next design to evaluate.
We use the Gaussian process (GP) as the surrogate model.
AF $\alpha(\bm{z})$ uses the surrogate models' posterior distribution to quantify the value of evaluating a new design.
As the inputs to BO are binary values in our case, optimizing AF is more difficult as gradient-based methods can not be directly applied.

Following Daulton et al.~\cite{DaultonWEBOB22}, we reparameterize the optimization problem by introducing a discrete probability distribution $p(\bm{Z} | \bm{\theta})$ parameterized by continuous parameter vector $\bm{\theta}$ over a random variable $\bm{Z}$ sampled from Bernoulli distribution.
Alg.~\ref{alg} illustrates the process.

\begin{algorithm}[t]
\caption{BO with Reparameterization}
\small
\begin{algorithmic}[1]
\State Input: objective $f : \Zstroke \to \mathbb{R}$
\State Initialize $\mathcal{D}_0 \gets \emptyset$, $\textbf{GP}_0 \gets \textbf{GP}(\textbf{0}, k)$
\For{$n = 1$ \textbf{to} $N_{\text{iterations}}$}
    \State $(\bm{\theta}_n) \gets \arg\max_{( \bm{\theta}) \in \Theta} \mathbb{E}_{\bm{Z} \sim p(\bm{Z}|\bm{\theta})} [\alpha(\bm{Z})]$
    \State Sample $\mathbf{z}_n \sim p(\bm{Z}|\bm{\theta}_n)$
    \State Evaluate $f(\mathbf{z}_n)$
    \State $\mathcal{D}_n \gets \mathcal{D}_{n-1} \cup \{(\bm{z}_n, f(\bm{z}_n))\}$
    \State Update posterior $\textbf{GP}_n$ given $\mathcal{D}_n$
\EndFor
\end{algorithmic}
\label{alg}
\end{algorithm}

Then, we can design the probabilistic objective (PO):
\begin{equation}
    \small
    \mathbb{E}_{\bm{Z}\sim p(\bm{Z}|{\bm{\theta}})}[\alpha(\bm{Z})].\nonumber
\end{equation}
We can optimize $\bm{\theta}$ over a continuous space to maximize PO. 
Since $p(\bm{Z} | \bm{\theta})$ is a discrete distribution over the discrete space $\Zstroke$, AF is only evaluated for discrete designs and PO can be rewritten as:
\begin{equation}
    \small
    \mathbb{E}_{\bm{Z}\thicksim p(\bm{Z}|\bm{\theta})}[\alpha(\bm{Z})]=\sum_{\bm{z}\in\Zstroke}p(\bm{z}|\bm{\theta})\alpha(\bm{z}).\nonumber
\end{equation}
Note that the above probabilistic objective is differentiable w.r.t. $\bm{\theta}$ and $\alpha(\bm{z})$ is not differentiable w.r.t. $\bm{z}$.
The gradient of PO is:
\begin{equation}
    \small
    \nabla_{\bm{\theta}}\mathbb{E}_{\bm{Z}\sim p(\bm{Z}|\bm{\theta})}[\alpha(\bm{Z})] =\sum_{\bm{z}\in\Zstroke}\alpha(\bm{z})\nabla_{\bm{\theta}}p(\bm{z}|\bm{\theta}).
    \nonumber
\end{equation}
Therefore, gradient-based methods can be used to optimize the probabilistic objective (Line 4 in Alg.~\ref{alg}). 
In practice, we use Monte Carlo sampling to estimate PO and its gradient to reduce the  cost~\cite{DaultonWEBOB22}.

\vspace{5pt}
\noindent\textbf{Router.} 
The goal of router $\mathcal{G}$ is to direct the input to the appropriate relation adapter.
To be precise, $\mathbf{q}^{k}_{i}$ and $\mathbf{p}^{k}_{i}$ for the $i$-th token are fused through router:
\begin{equation}
\boldsymbol{\beta} = \text{Softmax}\big([ \mathbf{q}^{k}_{i}, \mathbf{p}^{k}_{i} ]\mathbf{W}_{\text{gate}} + \mathbf{b}_{\text{gate}}\big),\,\,\,\,\mathbf{\tilde{h}}_{i}^{k} = \beta_1 \cdot \mathbf{q}^{k}_{i} + \beta_2 \cdot \mathbf{p}^{k}_{i}\nonumber
\end{equation}
where $\mathbf{W}_{\text{gate}}$ and $\mathbf{b}_{\text{gate}}$ are learnable parameters.
$\beta_1$ and $\beta_2$ are gating values in the first and second dimensions of $\boldsymbol{\beta} \in \mathbb{R}^2$, respectively.
$[*,*]$ indicates the concatenation operation.
$\mathbf{\tilde{h}}_{i}^{k}$ is passed to the recommendation adapter in $k-$th DALayer to produce $\mathbf{h}^k_i$.

\subsection{Putting All Together}

The overall objective is the combination of several adaption losses:
\begin{equation}
    \mathcal{L} = \mathcal{L}_{\text{seq}} + \lambda_1 \mathcal{L}_{\text{align}} + \lambda_2 \mathcal{L}_{\text{cl}} + \lambda_3 \mathcal{L}_{\text{cf}},\nonumber
\end{equation}
where $\lambda_1$, $\lambda_2$ and $\lambda_3$ are task weights.

The overall workflow of \ours is as follows: First, we apply Bayesian optimization on a subset of the data ($\text{Data}_{\text{BO}}$) to search the suitable structure of each DALayer. 
Then, we train \ours on the full dataset for sequential recommendation.

%% file: tex/experiment.tex

\section{Experiment}
\label{sec:exp}

\subsection{Experimental Settings}

\subsubsection{Datasets}

\begin{table}
    \centering
    \caption{Statistics of the datasets after preprocessing. ``Avg. Items'' denotes the average number of items in sequences.}
    \label{tab:data}
    \vspace{-10pt}
    \scalebox{0.95}{
\begin{tabular}{l|cccc}
    \toprule
    \textbf{Datasets}    & \textbf{\#Users} & \textbf{\#Items} & \textbf{\#Inter.} & \textbf{Avg. Items} \\ \hline
    \textbf{Arts}        & 56,083           & 22,685           & 364,356            & 8.50           \\
    \textbf{Scientific}  & 10,973           & 5,193            & 54,183             & 6.94           \\
    \textbf{Instruments} & 27,518           & 10,535           & 171,108            & 8.22           \\
    \textbf{Pantry}      & 14,178           & 4,961            & 102,661            & 9.24           \\
    \textbf{Games}       & 55,126           & 17,277           & 356,954            & 8.48    \\
    \toprule
\end{tabular}
    }
    \vspace{-5pt}
\end{table}

We use five sub-categories of Amazon datasets\footnote{\url{https://cseweb.ucsd.edu/~jmcauley/datasets/amazon_v2}} in our experiments, including ``Arts, Crafts, and Sewing'' (Arts), ``Industrial and Scientific'' (Scientific), ``Musical Instruments'' (Instruments), ``Prime Pantry'' (Pantry) and ``Video Games'' (Games).
We filter items without titles and items with fewer than four interactions.
Tab.~\ref{tab:data} presents the statistics of the data. 
We randomly selected $b\%$ of the data as $\text{Data}_{\text{BO}}$ for Bayesian optimization~\cite{LiuSY19, CaiZH19}\footnote{$b\%$ used for each dataset is listed in Appendix~\ref{sec:app_training}.}.

\begin{table*}[t]
    \centering
    \caption{Performance comparison of different methods. The best performance is highlighted in bold, while the second-best performance is underlined. The last column indicates the improvements over the best baseline models. All improvements are significant at $p < 0.01$ compared to the best baseline under the paired t-test.}
    \label{tab:performance}
    \vspace{-10pt}
    \scalebox{0.95}{
        \begin{tabular}{cccccccccccc}
        \toprule   
        \multicolumn{1}{l}{}         & \multicolumn{1}{l}{} & \multicolumn{4}{c}{\textbf{Non-LLM-based SRS}}                             & \multicolumn{5}{c}{\textbf{LLM-based SRS}}                                                                  & \textbf{}        \\ 
         \cmidrule(r){3-6} \cmidrule(r){7-11}
        \textbf{Dataset}             & \textbf{Metric}      & \textbf{GRU4Rec} & \textbf{SASRec} & \textbf{BERT4Rec} & \textbf{FDSA} & \textbf{P5}  & \textbf{UniSRec} & \textbf{RecFormer} & \textbf{LLMRec} & \textbf{\ours} & \textbf{Improv.} \\ \hline
        \multirow{3}{*}{Arts}        & NDCG@10              & 0.1118           & 0.1236          & 0.0950            & 0.1254        & {\ul 0.1286} & 0.0694           & 0.1251             & 0.0792          & \textbf{0.1400}               & 8.86\%           \\
                                     & Recall@10            & 0.1365           & {\ul 0.1619}    & 0.1255            & 0.1569        & 0.1482       & 0.1263           & 0.1580             & 0.1054          & \textbf{0.1767}               & 9.15\%           \\
                                     & MRR                  & 0.1082           & 0.1164          & 0.0890            & 0.1198        & {\ul 0.1234} & 0.0566           & 0.1198             & 0.0744          & \textbf{0.1340}               & 8.57\%           \\ \hline
        \multirow{3}{*}{Scientific}  & NDCG@10              & 0.0799           & 0.1001          & 0.0704            & 0.1018        & 0.1031       & 0.0656           & {\ul 0.1040}       & 0.0520          & \textbf{0.1162}               & 11.66\%          \\
                                     & Recall@10            & 0.1048           & 0.1430          & 0.0995            & 0.1337        & 0.1257       & 0.1226           & {\ul 0.1468}       & 0.0745          & \textbf{0.1529}               & 4.15\%           \\
                                     & MRR                  & 0.0771           & 0.0919          & 0.0654            & 0.0962        & {\ul 0.0977} & 0.0535           & 0.0966             & 0.0495          & \textbf{0.1105}               & 13.08\%          \\ \hline
        \multirow{3}{*}{Instruments} & NDCG@10              & 0.0797           & 0.0866          & 0.0703            & {\ul 0.0889}  & 0.0847       & 0.0700           & 0.0821             & 0.0486          & \textbf{0.0966}               & 8.61\%           \\
                                     & Recall@10            & 0.1047           & 0.1175          & 0.0906            & 0.1157        & 0.1056       & {\ul 0.1210}     & 0.1030             & 0.0689          & \textbf{0.1283}               & 6.00\%           \\
                                     & MRR                  & 0.0768           & 0.0825          & 0.0682            & {\ul 0.0857}  & 0.0801       & 0.0591           & 0.0803             & 0.0470          & \textbf{0.0927}               & 8.20\%           \\ \hline
        \multirow{3}{*}{Pantry}      & NDCG@10              & 0.0453           & 0.0570          & 0.0373            & {\ul 0.0618}  & 0.0409       & 0.0345           & 0.0566             & 0.0251          & \textbf{0.0689}               & 11.41\%          \\
                                     & Recall@10            & 0.0673           & 0.0887          & 0.0552            & 0.0848        & 0.0529       & 0.0731           & {\ul 0.0888}       & 0.0392          & \textbf{0.1036}               & 16.70\%          \\
                                     & MRR                  & 0.0435           & 0.0524          & 0.0363            & {\ul 0.0597}  & 0.0389       & 0.0278           & 0.0513             & 0.0251          & \textbf{0.0638}               & 6.84\%           \\ \hline
        \multirow{3}{*}{Games}       & NDCG@10              & 0.0743           & 0.0802          & 0.0604            & {\ul 0.0860}  & 0.0568       & 0.0489           & 0.0675             & 0.0339          & \textbf{0.0895}               & 4.05\%           \\
                                     & Recall@10            & 0.1142           & {\ul 0.1322}    & 0.0914            & 0.1273        & 0.0729       & 0.1020           & 0.1027             & 0.0534          & \textbf{0.1358}               & 2.73\%           \\
                                     & MRR                  & 0.0697           & 0.0728          & 0.0576            & {\ul 0.0810}  & 0.0535       & 0.0399           & 0.0513             & 0.0325          & \textbf{0.0839}               & 3.55\%           \\ \toprule   
        \end{tabular}
}
\end{table*}

\subsubsection{Evaluation}
We apply leave-one-out for evaluating SRS~\cite{KangM18}: the most recent item in the interaction sequence is used for testing, the second most recent item is for validation, and the remaining items are for training.
We adopt three popular evaluation metrics for SRS: Mean Reciprocal Rank (MRR), Recall@N, and Normalized Discounted Cumulative Gain (NDCG@N), where N is set to 10. 
We rank the ground-truth item of each sequence among all items and report the average scores across all sequences on the test set.

\subsubsection{Baselines}

We compare \ours with various baselines\footnote{Appendix~\ref{sec:app_baseline} provides the descriptions of each baseline.}, including (1) None-LLM-based SRS: 
GRU4Rec~\cite{HidasiKBT15}, SASRec~\cite{KangM18}, BERT4Rec~\cite{sun2019bert4rec} and  FDSA~\cite{zhang2019feature}.
(2) LLM-based SRS: P5\footnote{\url{https://github.com/jeykigung/P5}}~\cite{Geng0FGZ22}, UniSRec\footnote{\url{https://github.com/RUCAIBox/UniSRec}}~\cite{hou2022towards},  RecFormer\footnote{\url{https://github.com/JiachengLi1995/RecFormer}}~\cite{li2023text} and LLMRec.
LLMRec is a variation of \ours that removes adapter blocks from \oursbase (Fig.~\ref{fig:base}) and freezes LLM during fine-tuning. 
It is optimized with $\mathcal{L}_{\text{seq}}$ (Sec.~\ref{sec:predict_next}).
For a fair comparison, we only consider methods that rank all items as baselines. 
Besides, we use the pre-trained UniSRec and RecFormer provided by their authors and continue to fine-tune them following their papers.

\subsubsection{Implementation}

We use RecBole\footnote{\url{https://github.com/RUCAIBox/RecBole}}~\cite{recbole} to implement GRU4Rec, SASRec, BERT4Rec and FDSA. For other methods, we use implementations provided by the authors.
We use the small version for P5.
Other parameters are set following default configurations.

We train \ours on a machine with four NVIDIA RTX 3090 GPUs using the AdamW optimizer~\cite{LoshchilovH19}. 
We employ the pre-trained OPT-125M\footnote{\url{https://huggingface.co/facebook/opt-125m}} as the backbone of \oursbase and \ours. 
By default, we use Neural Collaborative Filtering (NCF)~\cite{HeLZNHC17} as CF model in \ours\footnote{In Appendix~\ref{sec:app_exp_cf}, we will compare the results using different CF models.}. 
We set the number of epochs to 40 and choose $5 \times 10^{-5}$ as the peak learning rate. 
The warmup strategy adjusts the learning rate during training, with the warmup stage set to the first 6\% of all iterations. 
The maximum item ``sentence'' length is set to 30.
The bottleneck size of the adapter block is 64.
For contrastive learning (Sec.~\ref{sec:cl}), within a sequence, we select one item ($n_s=1$) as the positive sample and choose five negative samples ($n_{cl}=5$) for that item by default. 
The temperature parameter $\tau$ for contrastive learning is set to 1.0.
For Bayesian optimization (Sec.~\ref{sec:bo}), we run 30 iterations to search for suitable structures of all DALayers. 
By default, we replace every LLM layer with DALayer (i.e., replacement frequency $r=1$).
The impact of these parameters, including $n_s$, $n_{cl}$, $r$, and the bottleneck size, is further analyzed in Sec.~\ref{sec:exp_para}.
We conduct a grid search to find the best hyper-parameters on different datasets\footnote{We report the best hyper-parameters on different datasets in Tab.~\ref{tab:training} of Appendix~\ref{sec:app_training}.}.

\subsection{Overall Performance}

Tab.~\ref{tab:performance} reports the overall performance of all methods\footnote{A comparison of running time of LLM-based methods is provided in Appendix~\ref{sec:app_runing time}.}.
Based on the results, we have the following observations:
\begin{itemize}[leftmargin=12pt,topsep=1pt,itemsep=0.5pt]
    \item Non-LLM-based methods SASRec and FDSA perform consistently well across different datasets. In some cases, their performance is close to or even better than LLM-based methods P5 and RecFormer.

    \item BERT4Rec performs poorly on several datasets, which is consistent with the observation mentioned in \cite{PetrovM22a}. As mentioned by Petrov and Macdonald, it is easy for BERT4Rec to get underfitted~\cite{PetrovM22a}, which may be the reason for its performance in our experiments.
        
    \item LLM-based methods P5 and RecFormer perform similarly to some extent. They perform well on  Arts and Scientific but do not exceed SASRec and FDSA on Instruments and Games. We speculate that the reason is that, in Instruments and Games datasets, user behaviors are more inclined toward instant interests and impulsive buying. P5 and RecFormer might not capture these short-term behavioral changes as effectively as traditional models. 

    \item UniSRec and LLMRec perform worse than P5 and RecFormer. As UniSRec, which is designed for cross-domain recommendation, is pre-trained on other sub-categories of Amazon datasets and then deployed on the five datasets used in our experiments, the discrepancy between the pre-trained data and recommendation data may lead to its mediocre performance. LLMRec is the degraded version of \ours. It does not use adapters and freezes LLM during fine-tuning. Therefore, it does not have the merits of \ours and performs much poorer than other other LLM-based methods. 
    
    \item \ours achieves the best performance across all datasets on all three metrics. Compared to the best baselines, \ours shows an average improvement of $8.92\%$ on NDCG@10, $7.75\%$ on Recall@10, and $8.05\%$ on MRR. The overall results demonstrate the effectiveness of \ours for sequential recommendation.
    
\end{itemize}

\subsection{Contribution of Each Part in \ours}
\label{sec:contribution}

\begin{figure*}[!th]
\centering
\begin{minipage}{.47\textwidth}
  \centering
  \includegraphics[width=1\columnwidth]{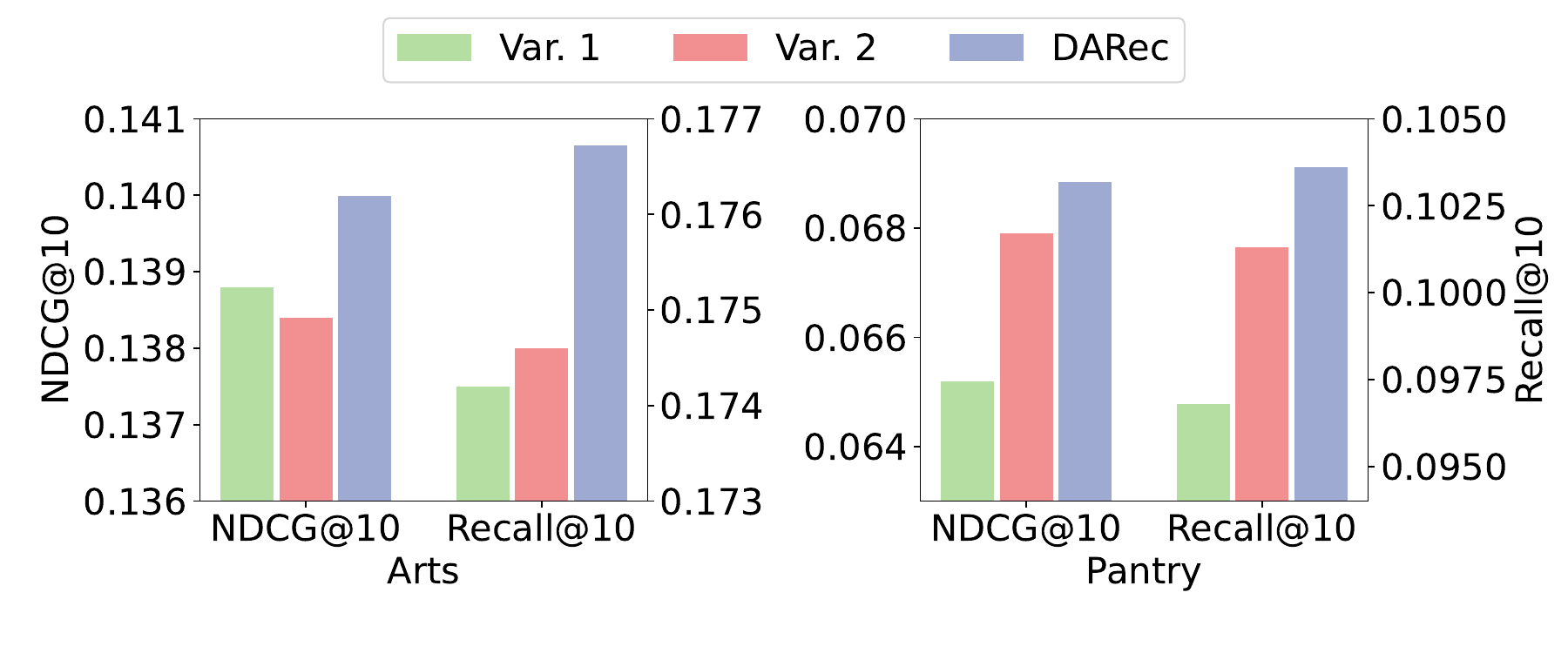}
  \vspace{-15pt}
  \caption{Comparison of \ours and its variations.}
  \label{fig:ablation}
\end{minipage}%
\hspace{0.03\textwidth} 
\begin{minipage}{.47\textwidth}
    \centering
  \includegraphics[width=1\columnwidth]{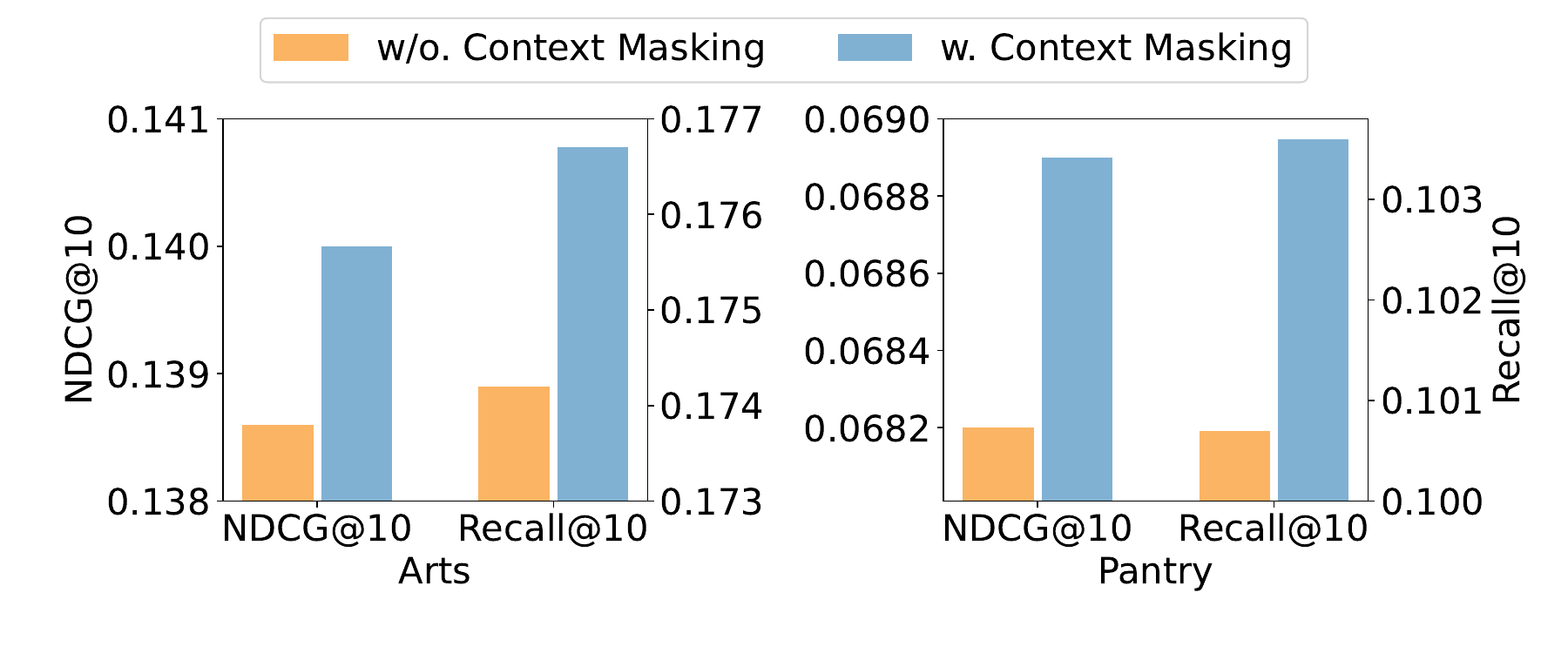}
  \vspace{-15pt}
  \caption{Effects of Using Context Masking.}
  \label{fig:result_Mask}
\end{minipage}
\end{figure*}

Firstly, we compare \ours and its two variations in Fig.~\ref{fig:ablation} to investigate the impacts of (a) coarse-grained adaption, (b) additionally incorporating intra-item relation and collaborative knowledge, and (c) using dynamic adaption: 
\begin{enumerate}[leftmargin=15pt,topsep=1pt,itemsep=0.5pt]

    \item \textbf{Variation 1}: It is \oursbase tuned with coarse-grained adaption tasks illustrated in Sec.~\ref{sec:adapt}. 

    \item \textbf{Variation 2}: It is \ours tuned with coarse-grained adaption (Sec.~\ref{sec:adapt}), context masking and collaborative filtering injection (Sec.~\ref{sec:enhance}) (i.e., remove dynamic adaption from \ours and uses serial adapter for all adapters).
\end{enumerate}

Due to the page limit, we only provide results of NDCG@10 and Recall@10 on Arts and Pantry.
From Fig.~\ref{fig:ablation}, we can observe that:
\begin{itemize}[leftmargin=12pt,topsep=1pt,itemsep=0.5pt]
    \item Variation 1 outperforms LLMRec in Tab.~\ref{tab:performance}, showing that injecting adapters into LLM and fine-tuning adapters with our proposed coarse-grained adaption tasks (semantic alignment and distinguishing hard samples) for sequential recommendation can improve the performance of LLM-based SRS.  

    \item Variation 2 exceeds variation 1 on Pantry, but it lags behind variation 1 on Arts w.r.t. NDCG@10. The results show that including context masking and collaborative filtering injection can enhance the performance, but the improvement can be offset by the fixed adapter design.

    \item \ours surpasses all its variations for all cases. Particularly, on Arts where variation 2 falls behind variation 1, \ours (i.e., enhance variation 2 with Bayesian optimization) exceeds variation 1, showing that using Bayesian optimization to dynamically search for the suitable adapter architecture indeed enhances the quality of sequential recommendation.
\end{itemize}

However, the above analysis is insufficient for verifying the effectiveness of context masking, as both intra-item relation (captured by context masking) and collaborative knowledge are considered in variation 2 and \ours.
We further compare the results of \ours and using two relation adapters that both adopt causal masking on Arts and Pantry.
From the results in Fig.~\ref{fig:result_Mask}, we can see that context masking indeed enhances the performance of sequential recommendation compared to solely using causal masking.

We further explore the effects of using different CF models and different ways to train the CF model\footnote{The results and analysis are reported in Appendix~\ref{sec:app_exp_cf}.}.

Overall, we can conclude that each part in \ours contributes to its performance on sequential recommendation.

\subsection{Effects of Different Hyper-Parameters}
\label{sec:exp_para}

\begin{figure*}[!th]
\centering
\begin{minipage}{.72\textwidth}
  \centering
  \includegraphics[width=1\linewidth]{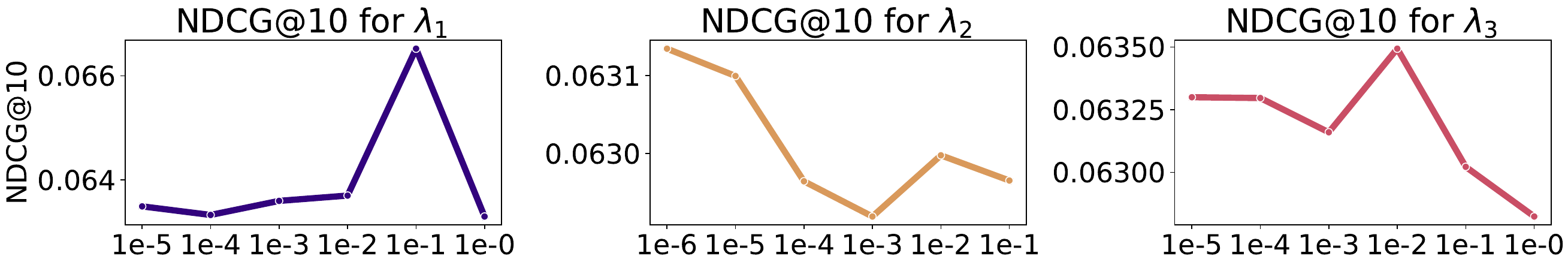}
  \vspace{-15pt}
  \captionof{figure}{Performance when varying ${\lambda}_1$, ${\lambda}_2$ and ${\lambda}_3$.\\}
  \label{fig:lambda_comparison}
\end{minipage}%
\hspace{0.01\textwidth} 
\begin{minipage}{.23\textwidth}
  \centering
  \vspace{12pt}
  \includegraphics[width=1\linewidth]{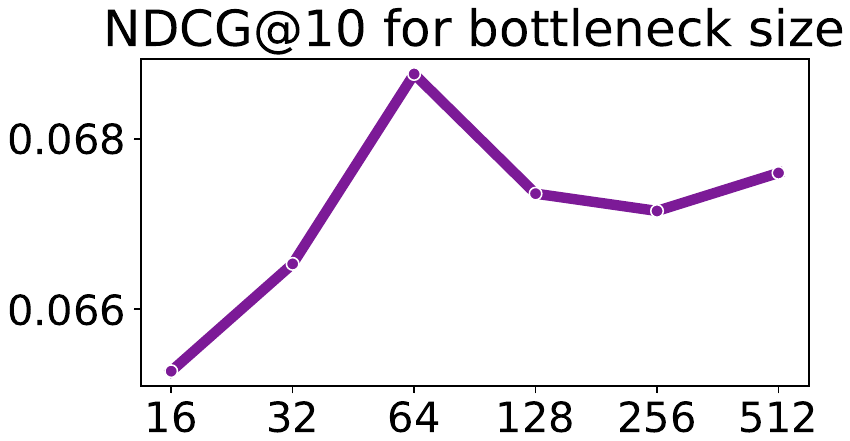}
  \vspace{-15pt}
  \captionof{figure}{Performance when varying bottleneck size.}
  \label{fig:size_comparison}
\end{minipage}
\end{figure*}

\begin{figure*}[!th]
\centering
\begin{minipage}{.67\textwidth}
  \centering
  \includegraphics[width=1\columnwidth]{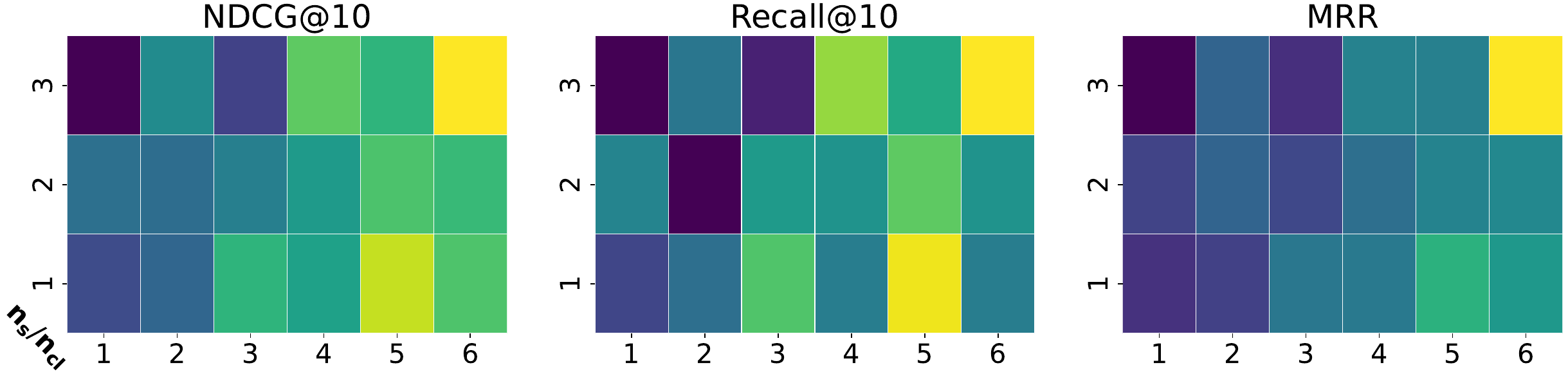}
  \vspace{-15pt}
  \captionof{figure}{Performance when varying $n_s$ and $n_{cl}$. Darker color indicates lower values.}
  \label{fig:heatmap}
\end{minipage}%
\hspace{0.05\textwidth} 
\begin{minipage}{.27\textwidth}
  \centering
  \includegraphics[width=1\linewidth]{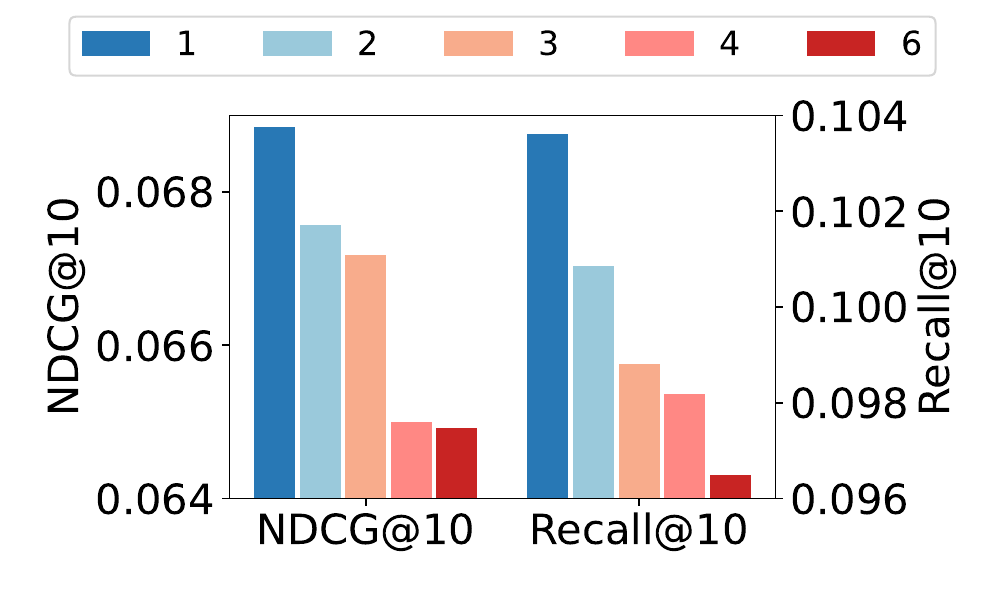}
  \vspace{-15pt}
  \captionof{figure}{Performance when varying replacement frequency $r$.}
  \label{fig:result_freq}
\end{minipage}
\end{figure*}

In this section, we investigate the effects of several hyper-parameters on \ours and provide analysis based on Pantry:
\begin{itemize}[leftmargin=12pt,topsep=1pt,itemsep=0.5pt]
    \item \textbf{Task Weights}: Fig.~\ref{fig:lambda_comparison} shows the impact on \ours when adjusting one $\lambda$ value while keeping the others fixed at 1. As depicted, the performance peaks around ${\lambda}_1=1e-1$, ${\lambda}_2=1e-6$ and ${\lambda}_3=1e-2$. We can also clearly see the performance peak and the performance lines do not vibrate. In other words, it is not difficult to tune task weights for \ours.
    
    \item \textbf{Bottleneck Size:}: As illustrated in Fig.~\ref{fig:size_comparison}, \ours achieves the best performance when the bottleneck size is 64. Besides, the performance trend when varying bottleneck size is clear, showing that tuning bottleneck size is not difficult in practice.
    
    \item \textbf{Sample Number:} Fig.~\ref{fig:heatmap} depicts the performance when using different values for $n_s$ and $n_{cl}$. It can be observed that, as $n_s$ and $n_{cl}$ increase, the performance generally gets improves. To balance the performance and the computation cost, we set $n_s=1$ and $n_{cl}=5$ by default.

    \item \textbf{Frequency of Replacement:} By default, the replacement frequency $r$ is set to 1. Fig.~\ref{fig:result_freq} illustrates the performance of \ours under varying replacement frequencies, with $r$ values of 1, 2, 3, 4, and 6. For instance, $r=2$ indicates that we replace LLM layer with DALayer every two layers. For unchanged layers, we use the original LLM layers and keep their parameters frozen. From the results, we can see that, as more LLM layers are replaced with DALayers, the performance of \ours increases, showing that our proposed method can effectively improve LLM-based sequential recommendation.
    
\end{itemize}

%% file: tex/related_work.tex

\section{Related Work}
\label{sec:related}

\subsection{Non-LLM-Based SRS}

Traditional SRS (non-LLM-based SRS) typically apply sequential pattern mining~\cite{YapLY12}, Markov chains based methods~\cite{HeFWM16, HeM16}, and factorization based methods~\cite{RendleFS10,ChengYLK13}.
More recent works adopt deep learning techniques to boost the performance of sequential recommendation.
Researchers have explored RNNs~\cite{HidasiKBT15,QuadranaKHC17}, CNNs~\cite{TuanP17,TangW18}, GNNs~\cite{WuT0WXT19,QiuHLY20}, Transformer~\cite{KangM18,RenLLZWDW20,sun2019bert4rec} and many other prevalent deep learning techniques in the sequential recommendation task and find they can reduce the cost of manual feature engineering and better capture the comprehensive sequential relations~\cite{WangHWCSO19}. 
Readers can refer to survey papers on SRS~\cite{FangZSG20,WangHWCSO19,WangZ0ZW022,QuadranaCJ18} for detailed descriptions of non-LLM-based SRS.

\subsection{LLM-Based SRS}

Emerging LLM-based SRS can be grouped into two paradigms: LLM-augmented SRS and LLM-centric SRS~\cite{HuXZFWHLTZ24}.

LLM-augmented SRS adopt LLM as the feature extractor to obtain item features that are further fed to SRS.
Harte et al.~\cite{HarteZLKJF23} show that initializing BERT4Rec~\cite{sun2019bert4rec} with embeddings obtained from LLM can significantly improve sequential recommendation.
LRD~\cite{YangMSALCZ24} is a LLM-based latent relation discovery framework.
It uses language knowledge to discover latent relations that can further enhance SRS.
SERALM underpins alignment training method to refine LLMs' generation using feedback from ID-based recommenders for better knowledge augmentation~\cite{RenCYLJCZM024}.

LLM-centric SRS leverage LLM as the SRS.
RecFormer~\cite{li2023text} describes items using item ``sentences''.
It is trained to understand the ``sentence'' sequence and retrieve the next ``sentence'' for making sequential recommendations.
E4SRec~\cite{li2023e4srec} takes ID sequences as inputs to LLM and ensures that the generated output falls within the candidate lists.
LLM-TRSR~\cite{ZhengCQZ024} focuses on modeling over-length user behavior sequences.
It segments behavior sequences and applies summarization techniques to form the inputs to LLM-based SRS.
SAID~\cite{HuXZFWHLTZ24} uses LLM to learn semantically aligned item ID embeddings that can be integrated with downstream SRS.
LLM4ISR~\cite{0001LQFWO24}, armed with prompt initialization, optimization, and selection modules, enables LLM to make intent-driven session recommendations.
Re2LLM~\cite{abs-2403-16427} guides LLM with self-reflection and a lightweight retrieval agent, enabling LLM to focus on specialized knowledge essential for more accurate sequential recommendations effectively and efficiently.
Chu et al.~\cite{abs-2405-02778} propose three prompting strategies to exploit temporal information within historical interactions for LLM-based SRS.

\subsection{PEFT for LLM}

Existing PEFT can be categorized into four groups:
(1) Additive PEFT maintains the pre-trained LLM unchanged and introduces only a small number of trainable parameters that are strategically positioned within the model architecture.
Representative works include serial adapter~\cite{HoulsbyGJMLGAG19}, parallel adapter~\cite{HeZMBN22} and prefix-tuning~\cite{LiL20}.
(2) Selective PEFT fine-tunes a subset of LLM parameters to enhance performance over downstream tasks. 
Diff pruning~\cite{GuoRK20} is a representative selective PEFT method.
(3) Reparameterized PEFT introduces additional low-rank trainable parameters during training, which are then integrated with the original LLM for inference.
LoRA (Low Rank Adaptation)~\cite{HuSWALWWC22} is the most widely recognized reparameterized PEFT technique.
(4) Hybrid PEFT combines the advantages of diverse PEFT approaches~\cite{MaoMHAM0YK22}.

%% file: tex/conclusion.tex

\section{Conclusion}
\label{sec:con}

In this paper, we propose \ours for enhancing LLM-based sequential recommendation beyond modeling inter-item relations. 
Built on top of coarse-grained adaption, \ours is further enhanced with context masking, collaborative knowledge injection and a dynamic adaption mechanism so that it can effectively handle sequential recommendation. 
In the future, we plan to try other PEFT methods in \ours (e.g., selective PEFT) and improve their flexibility in \ours.
Moreover, to understand the robustness of \ours, we will investigate the impact of using LLMs with different sizes as the backbone in \ours.

%% file: tex/appendix.tex

\appendix

\section{Generate Hard Samples}
\label{sec:app_hard_sample}

To construct hard samples, we adopt a pre-trained BERT model\footnote{\url{https://huggingface.co/google-bert/bert-base-uncased}}~\cite{DevlinCLT19} to encode item text embeddings and matrix factorization (MF)~\cite{KorenBV09} to produce item latent vectors.
First, we use the pre-trained BERT model to encode the item ``sentence'' of each item $j$ and use the average of representations of tokens in $j$'s item ``sentence'' as the item text embedding for $j$. 
Then, we perform matrix factorization~\cite{KorenBV09} over user-item interaction matrix to generate item latent vectors with the dimensionality of 1,000.

After that, for each item $j$, we construct two similar item sets: one contains the top 0.5\% most similar items to $j$ w.r.t. L2 distance between item text embeddings, and the other contains the top 0.5\% most similar items to $j$ w.r.t. L2 distance between item latent vectors.
The union of the two sets, excluding all items that co-occur with $j$ in any user historical interaction sequence $S$, is the hard item set $N_j$ of $j$.

\section{Demonstration of Two Masking Mechanisms for Semantic Alignment}
\label{sec:app_align}

\begin{figure}[H]
\centering
  \centering
  \includegraphics[width=1\columnwidth]{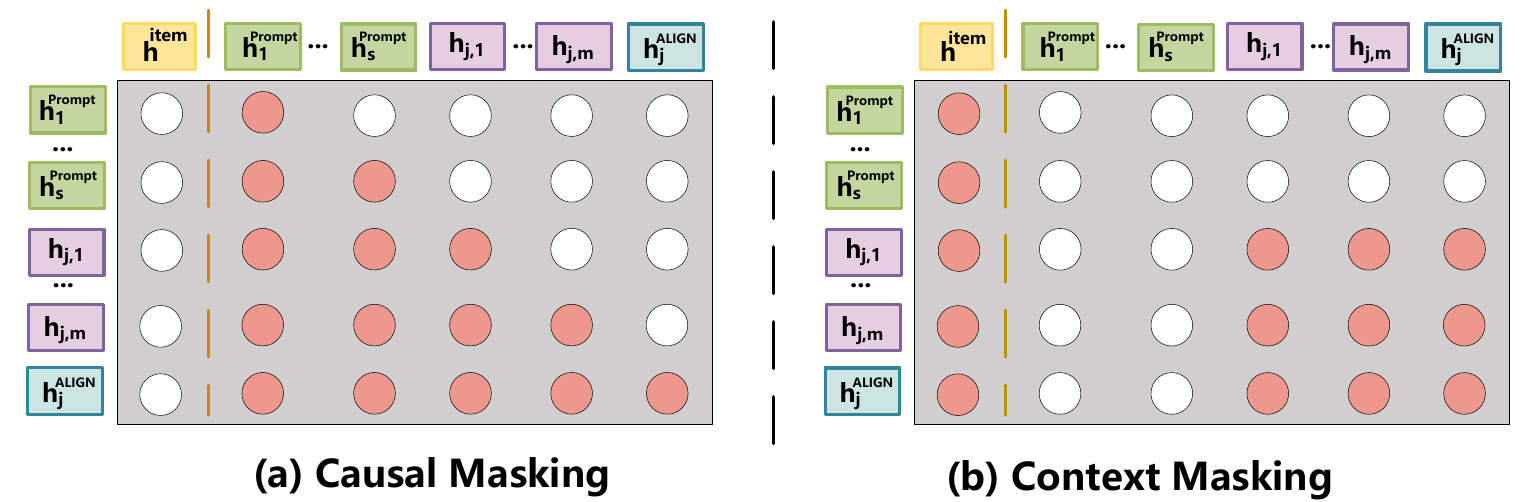}
  \vspace{-20pt}
  \captionof{figure}{Two Masking Mechanisms for Semantic Alignment.}
  \label{fig:mask_align}
\end{figure}

Fig.~\ref{fig:mask_align} demonstrates the two masking mechanisms for semantic alignment.

\section{Attention Calculation}
\label{sec:app_att}

For the $k$-th LLM layer, the hidden state $h_i^k$ of the $i$-th token is computed using causal masking and context masking (hidden states of all tokens are denoted by $\mathbf{H}^k$). These two masking mechanisms facilitate the computation of attention with hidden states of other tokens, resulting in two distinct outputs.

Initially, we derive query matrix $\mathbf{Q}^k$, key matrix $\mathbf{K}^k$, and value matrix $\mathbf{V}^k$ through linear transformations, formulated as follows:
\begin{equation}
\small
\begin{aligned}
    \mathbf{Q}^k &= \text{Linear}_{\mathbf{Q}^k}(\mathbf{H}^k) = \mathbf{H}^k \mathbf{W}_{\mathbf{Q}^k} + \mathbf{b}_{\mathbf{Q}^k} \\
    \mathbf{K}^k &= \text{Linear}_{\mathbf{K}^k}(\mathbf{H}^k) = \mathbf{H}^k \mathbf{W}_{\mathbf{K}^k} + \mathbf{b}_{\mathbf{K}^k} \\
    \mathbf{V}^k &= \text{Linear}_{V^k}(\mathbf{H}^k) = \mathbf{H}^k \mathbf{W}_{\mathbf{V}^k} + \mathbf{b}_{\mathbf{V}^k}\nonumber
\end{aligned}
\end{equation}
where $\mathbf{W}$ and $\mathbf{b}$ are learnable weights.

Subsequently, the attention weights are computed using causal masking and context masking, and the outputs are:
\begin{equation}
\small
\begin{aligned}
    \text{Linear}(\mathbf{H}) &= \mathbf{W}_{\text{Linear}}\mathbf{H} + \mathbf{B}_{\text{Linear}}\\
    \text{Self-Attention}(\mathbf{H}^k,\textbf{MASK}) &= \text{Linear}\Big(\text{Softmax}\left(\frac{\mathbf{Q}^k (\mathbf{K}^k)^T}{\sqrt{d_{LLM}}} \odot \textbf{MASK}\right) \mathbf{V}^k \Big) \nonumber
\end{aligned}
\end{equation}
where $d_{LLM}$ is the dimensionality of hidden states, $\odot$ indicates element-wise product, and $\mathbf{W}$ and $\mathbf{B}$ are learnable weights.
Note that, in Sec.~\ref{sec:bo}, we use $\text{Self-Attention}(\mathbf{h},\textbf{MASK})$ for illustration where $\mathbf{h}$ is the hidden state of a token, but its definition is similar. 

\section{Demonstration of Serial Adapter and Parallel Adapter}
\label{sec:app_adapter}

\begin{figure}[H]
\centering
  \centering
  \includegraphics[width=0.9\columnwidth]{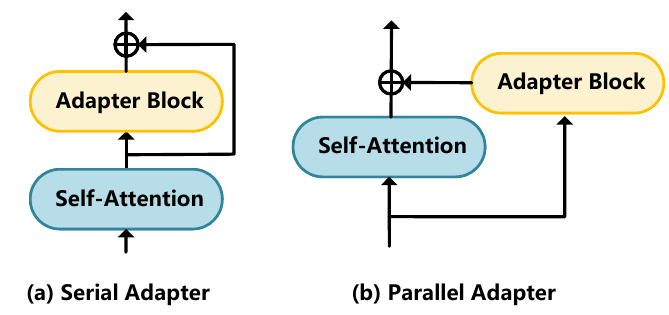}
  \vspace{-10pt}
  \captionof{figure}{Two Masking Mechanisms for Semantic Alignment.}
  \label{fig:two_adapter}
  \vspace{-10pt}
\end{figure}

Fig.~\ref{fig:two_adapter} visualizes the designs of serial adapter and parallel adapter.

\section{Hyper-Parameters for Different Datasets}
\label{sec:app_training}

We perform a grid search to find optimal hyper-parameters for \ours on different datasets.
We report them in Tab.~\ref{tab:training}.

\begin{table}[!th]
    \centering
  \vspace{-5pt}
    \caption{Hyper-Parameters for different datasets.}
    \vspace{-10pt}
    \scalebox{0.9}{
\begin{tabular}{c|ccccc}
\toprule
\textbf{Parameter}     & Arts  & Scientific & Instruments & Pantry & Games \\ \hline
$n_s$                  & \multicolumn{5}{c}{1}                             \\ \hline
$n_{cl}$               & \multicolumn{5}{c}{5}                             \\ \hline
Epochs                 & \multicolumn{5}{c}{40}                            \\ \hline
Learning Rate          & \multicolumn{5}{c}{5e-5}                          \\ \hline
LR Scheduler           & \multicolumn{5}{c}{Cosine Scheduler}              \\ \hline
Bottleneck Size        & \multicolumn{5}{c}{64}                            \\ \hline
Batch Size             & 5     & 6          & 7           & 8      & 8     \\ \hline
$b\%$             & 0.04  & 0.2        & 0.1         & 0.2    & 0.06  \\ \hline
${\lambda}_1$          & 0.1   & 0.2        & 0.1         & 0.2    & 0.05  \\ \hline
${\lambda}_2$          & 1e-6  & 1e-3       & 1e-3        & 1e-6   & 1e-3  \\ \hline
${\lambda}_3$          & 0.001 & 0.1        & 0.05        & 0.1    & 0.01  \\ \toprule
\end{tabular}
}
  \vspace{-15pt}
    \label{tab:training}
\end{table}

\hide{

\section{Bayesian Optimization with Reparameterization}
\label{sec:app_BO}

Alg.~\ref{alg} illustrates how \ours adopts Bayesian optimization with reparameterization to search for suitable architecture for DAMoE.

\begin{algorithm}[H]
\caption{BO with Reparameterization}
\begin{algorithmic}[1]
\State Input: objective $f : \Zstroke \to \mathbb{R}$
\State Initialize $\mathcal{D}_0 \gets \emptyset$, $\textbf{GP}_0 \gets \textbf{GP}(\textbf{0}, k)$
\For{$n = 1$ \textbf{to} $N_{\text{iterations}}$}
    \State $(\bm{\theta}_n) \gets \arg\max_{( \bm{\theta}) \in \Theta} \mathbb{E}_{\bm{Z} \sim p(\bm{Z}|\bm{\theta})} [\alpha(\bm{Z})]$
    \State Sample $\mathbf{z}_n \sim p(\bm{Z}|\bm{\theta}_n)$
    \State Evaluate $f(\mathbf{z}_n)$
    \State $\mathcal{D}_n \gets \mathcal{D}_{n-1} \cup \{(\bm{z}_n, f(\bm{z}_n))\}$
    \State Update posterior $\textbf{GP}_n$ given $\mathcal{D}_n$
\EndFor
\end{algorithmic}
\label{alg}
\end{algorithm}
}

\section{Baselines}
\label{sec:app_baseline}

We use the following baselines in our experiments:

\vspace{3pt}
\noindent\textbf{(1) Traditional SRS:}
\begin{itemize}[leftmargin=12pt,topsep=1pt,itemsep=0.5pt]
  \item \textbf{GRU4Rec}~\cite{HidasiKBT15} leverages gated recurrent units (GRU) to model user behavior sequences, effectively capturing historical interactions to construct comprehensive user profiles. Each user’s sequence is treated as a session.
  \item \textbf{SASRec}~\cite{KangM18} employs a directional self-attention mechanism to process item embeddings sequentially, aggregating past user interactions to predict future interests.
  \item \textbf{BERT4Rec}~\cite{sun2019bert4rec} utilizes a bidirectional self-attention model with a cloze task to model user behavior sequences.
  \item \textbf{FDSA}~\cite{zhang2019feature} integrates two self-attention blocks that process both item-level and feature-level sequences, merging their outputs to enhance the accuracy of next-item predictions.
  
\end{itemize}

\vspace{3pt}
\noindent\textbf{(2) LLM-based SRS:}
\begin{itemize}[leftmargin=12pt,topsep=1pt,itemsep=0.5pt]
  \item \textbf{P5}\footnote{\url{https://github.com/jeykigung/P5}}~\cite{Geng0FGZ22} is a unified framework that transforms all recommendation data into natural language sequences, facilitating personalized recommendations.
  \item \textbf{UniSRec}\footnote{\url{https://github.com/RUCAIBox/UniSRec}}~\cite{hou2022towards} embeds item text using a lightweight architecture based on parametric whitening and an MoE-enhanced adaptor to learn universal item representations. 
  \item \textbf{RecFormer}\footnote{\url{https://github.com/JiachengLi1995/RecFormer}}~\cite{li2023text} relies on an encoder-only architecture, learning user preferences and item features from natural language.
  \item \textbf{LLMRec}: It removes adapter blocks from \oursbase (Fig.~\ref{fig:base}) and freezes LLM during fine-tuning. 
It is optimized with $\mathcal{L}_{\text{seq}}$ (Sec.~\ref{sec:predict_next}).
\end{itemize}

\section{Effects of Using Different CF Models}
\label{sec:app_exp_cf}

\begin{figure}[H]
  \centering
  \vspace{-5pt}
  \includegraphics[width=1\columnwidth]{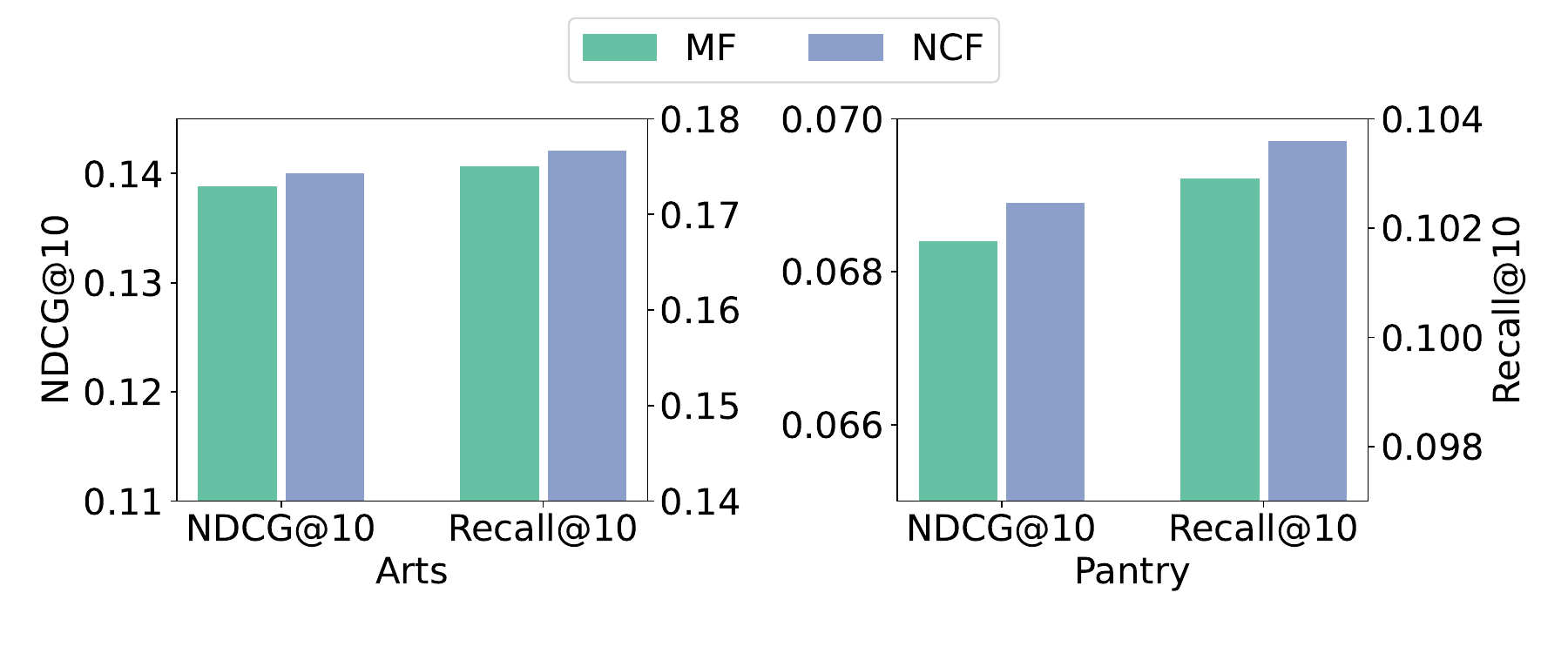}
  \vspace{-20pt}
  \caption{Effects of different CF models.}
  \vspace{-5pt}
  \label{fig:result_CF}
\end{figure}

Fig.~\ref{fig:result_CF} compares the performance of \ours using different CF models.
We apply two prevalent CF methods MF~\cite{KorenBV09} and NCF~\cite{HeLZNHC17}.
We can see that, using NCF generally brings better results than using MF, showing that introducing more sophisticated CF models can enhance \ours.

\begin{figure}[H]
  \centering
  \vspace{-5pt}
  \includegraphics[width=1\columnwidth]{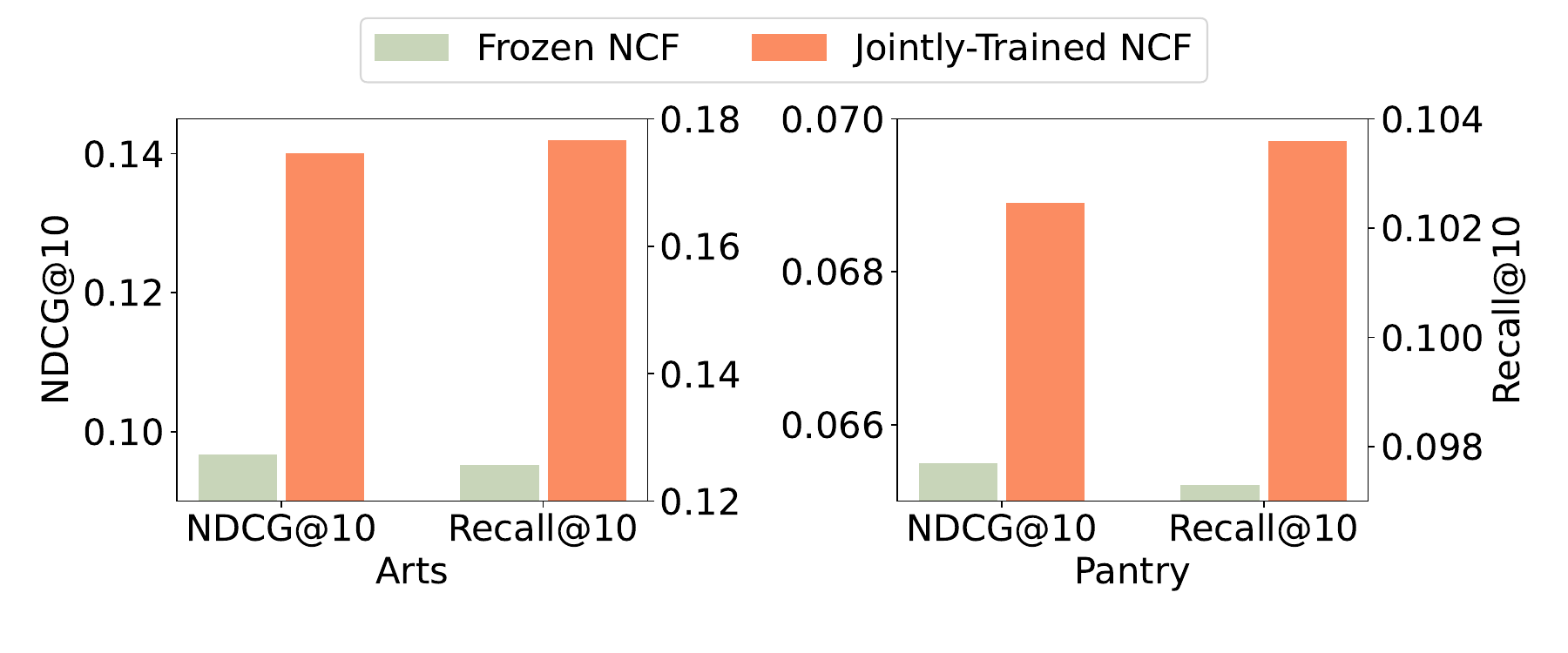}
  \vspace{-20pt}
  \caption{Effects of Frozen vs. Jointly-Trained NCF.}
  \vspace{-5pt}
  \label{fig:CF_fro}
\end{figure}

By default, we jointly train CF model and \ours.
We also investigate the case where the CF model is pre-trained and then frozen in collaborative information injection.
Fig.~\ref{fig:CF_fro} reports the results on Arts and Pantry when using frozen NCF and jointly-train NCF (i.e., default \ours).
The results show that jointly training CF model and \ours can improve sequential recommendation.

\section{A Comparison of Running Time}
\label{sec:app_runing time}

\begin{table}[!th]
    \centering
    \caption{A Comparison of Running Time.}
    \vspace{-10pt}
    \scalebox{0.9} {
\begin{tabular}{ccccc}
    \toprule  
                & \multicolumn{2}{c}{\textbf{Arts}}                                          & \multicolumn{2}{c}{\textbf{Pantry}}                   \\
    \cmidrule(lr){2-3} \cmidrule(lr){4-5}
\textbf{Method} & \textbf{\begin{tabular}[c]{@{}c@{}}Train\\ time (h)\end{tabular}} & \multicolumn{1}{c|}{\textbf{\begin{tabular}[c]{@{}c@{}}Inference\\ time (s)\end{tabular}}} & \textbf{\begin{tabular}[c]{@{}c@{}}Train\\ time (h)\end{tabular}} & \textbf{\begin{tabular}[c]{@{}c@{}}Inference\\ time (s)\end{tabular}} \\ \hline
RecFormer       & 43.46                                                             & \multicolumn{1}{c|}{796}                                                                   & 4.32                                                              & 141                                                                   \\ \hline
LLM-Rec         & 1.56                                                              & \multicolumn{1}{c|}{86}                                                                    & 0.34                                                              & 20                                                                    \\ \hline
DARec           & 12.08                                                             & \multicolumn{1}{c|}{575}                                                                   & 2.25                                                              & 142                                                                                      
    \\ \toprule  
\end{tabular}
    }

    \label{tab:ex_time}
\end{table}

Tab.~\ref{tab:ex_time} reports the training and test (inference) time of three LLM-based methods RecFormer, LLM-Rec and \ours on Arts and Pantry.
From the results, we can see that LLM-Rec uses the least running time as it does not use adapters and freeze LLM.
However, its performance is much worse than RecFormer and \ours (Tab.~\ref{tab:performance}) as it loses the merits of adaption.
\ours requires less running time than RecFormer, showing that our proposed method can achieve both effectiveness and efficiency compared to existing LLM-based methods.